# High Seebeck coefficient and ultra-low lattice thermal conductivity in $Cs_2InAgCl_6$

Enamul Haque and M. Anwar Hossain*

Department of Physics, Mawlana Bhashani Science and Technology University, Santosh, Tangail-1902, Bangladesh

Email: anwar647@mbstu.ac.bd, enamul.phy15@yahoo.com

**Abstract**

The elastic, electronic and thermoelectric properties of indium-based double-perovskite halide, $Cs_2InAgCl_6$ have been studied by first principles study. The $Cs_2InAgCl_6$ is found to be elastically stable, ductile, anisotropic and relatively low hard material. The calculated direct bandgap 3.67 eV by TB-mBJ functional fairly agrees with the experimentally measured value 3.3 eV but PBE functional underestimates the bandgap by 1.483 eV. The relaxation time and lattice thermal conductivity have been calculated by using relaxation time approximation (RTA) within the supercell approach. The lattice thermal conductivity ($\kappa_l$) is quite low (0.2 $Wm^{-1}K^{-1}$). The quite low phonon group velocity in the large weighted phase space, and high anharmonicity (large phonon scattering) are responsible for small $\kappa_l$. The room temperature Seebeck coefficient is 199 $\mu VK^{-1}$. Such high Seebeck coefficient arises from the combination of the flat conduction band and large bandgap. We obtain power factors at 300K by using PBE and TB-mBJ potentials are ~29 and ~31 $mWm^{-1}K^{-2}$, respectively and the corresponding thermoelectric figure of merit of $Cs_2BiAgCl_6$ are 0.71 and 0.72. However, the maximum ZT value obtained at 700K is ~0.74 by TB-mBJ potential. The obtained results implies that $Cs_2InAgCl_6$ is a promising material for thermoelectric device applications.

1. Introduction

The lead halides have gained huge interest because of their semiconducting behavior and suitable in perovskite solar cell applications. Lead-halides can efficiently absorb light in the visible range [1], have long diffusion length [2], as well as the photons emitted in the radiative recombination process are recyclable [2]. Due to the lack of structural stability and the increased of temperature and humidity when the cells (making with these halides) are exposed to light [3–7], the lead (Pb) has been successfully replaced by Bi and Ag [8–10]. A double-perovskite halide is formed due to the replacement of lead. Recently, it has been found that $Cs_2BiAgCl_6$ and $Cs_2BiAgBr_6$ are stable and have a highly tunable bandgap in the visible range [8–10]. For $Cs_2BiAgBr_6$, no phase transition was observed in 703K [8]. The optical measurement and band structure calculation reveal that both halides are indirect bandgap semiconductors [8–10]. Due to indirect bandgap, these are not suitable for thin film photovoltaic applications. Volonaki*et al*. have recently reported another lead free double perovskite halide $Cs_2InAgCl_6$ that have a highly tunable and direct bandgap [11]. The synthesized powders of $Cs_2AgInCl_6$ exhibit white coloration with the measured optical bandgap of 3.3 eV while the photoluminescence emission energy is only about 2.0 eV [12]. The parity forbidden transitions can explain such mystery of this lead free halide [12]. Recently, many experimental and theoretical studies have been reported on the electronic and optical properties of $Cs_2AgInCl_6$ to reveal the suitability of it for photovoltaic solar cell application [11–13]. Recent study on transport properties of lead free halides $Cs2BiAgX_6$ (*X*=Cl, Br) reveals that these materials have potential in thermoelectric device applications [14,15]. However, the suitability of $Cs_2AgInCl_6$ as a thermoelectric material is not studied yet. A good thermoelectric material should has high figure of merit and is defined as [16]

$ZT = \frac{S^2\sigma}{k_e+k_l} T$, where S is the thermopower, σ is the electrical conductivity and $\kappa_e + \kappa_l$ is sum of lattice and electronic thermal conductivity. A material having ZT ~1.0 is considered as good thermoelectric material. Such high thermoelectric figure of merit may be obtained when the power factor $S^2\sigma$ is high and thermal conductivity is low. Usually, semiconducting materials can exhibit good thermoelectric properties.

In this paper, we have presented first principles calculations to study the elastic, electronic, transport and lattice dynamical properties of $Cs_2InAgCl_6$ halide. Ultra-low lattice thermal conductivity in $Cs_2InAgCl_6$ halide arises from large weighted phase space and group velocity. The low value of total thermal conductivity makes $Cs_2InAgCl_6$ suitable for thermoelectric application.

## 2. Computational methods

The elastic and electronic properties were studied by using full potential linearized augmented plane wave (LAPW) as implemented in WIEN2k [17]. For good convergence, a plane wave cutoff of kinetic energy $RK_{max}$ =7.5 and $21 \times 21 \times 21$ k-point in Brillouin zone integration were chosen. The muffin tin radii 2.5 for Cs, Bi, Ag, and 2.15 for Cl, were used. The PBE-GGA functional [18,19] and modified TB-mBJ [20] functional were used in the electronic structure and transport properties calculations. Since the spin-orbit coupling has no significant effect on the electronic properties of $Cs_2InAgCl_6$ [11], we did not include spin-orbit coupling effect in our calculations. The energy and charge convergence criteria were set to $10^{-4} Ry$ and 0.001e, respectively. The transport properties were calculated by BoltzTraP code [21] by using $43 \times 43 \times 43$ k-point in WIEN2k to generate the required input files and the chemical potential was set to the zero temperature Fermi energy. The solution of semi-classical Boltzmann transport equation allows us to calculate the transport coefficients. The transport coefficients are defined in the Boltzmann transport theory [22–24] as

$$\sigma_{\alpha\beta}(T,\mu) = \frac{1}{V}\int \Sigma_{\alpha\beta}(\varepsilon)\left[-\frac{\partial f_\mu(T,\varepsilon)}{\partial \varepsilon}\right]d\varepsilon \quad (1)$$

$$S_{\alpha\beta}(T,\mu) = \frac{1}{eTV\sigma_{\alpha\beta}}\int \Sigma_{\alpha\beta}(\varepsilon)(\varepsilon-\mu)\left[-\frac{\partial f_\mu(T,\varepsilon)}{\partial \varepsilon}\right]d\varepsilon \quad (2)$$

$$K^e_{\alpha\beta}(T,\mu) = \frac{1}{e^2TV}\int \Sigma_{\alpha\beta}(\varepsilon)(\varepsilon-\mu)^2\left[-\frac{\partial f_\mu(T,\varepsilon)}{\partial \varepsilon}\right]d\varepsilon \quad (3)$$

where V is the volume of the unit cell, $\alpha$ and $\beta$ represent Cartesian indices, $\mu$ is the chemical potential and $f_\mu$ Fermi-Dirac function. The energy projected conductivity tensors are defined as

$$\Sigma_{\alpha\beta}(\varepsilon) = \frac{e^2}{N}\sum_{i,k}\tau_{i,k}v_\alpha(i,k)v_\beta(i,k)\frac{\delta(\varepsilon-\varepsilon_{i,k})}{d\varepsilon} \quad (4)$$

where N is the number of sampling k-points for BZ integration, $i$ is the index of band, $v$ and $\tau$ stand for electrons group velocity and relaxation time. In BoltzTraP transport coefficients calculation, the constant relaxation time approximation is used. To calculate lattice thermal conductivity, the supercell approach was used through creating total 412 displacements and choosing $16 \times 16 \times 16$ mesh for BZ integration in thePhono3py program [25]. The required forces for lattice thermal conductivity calculation were obtained by using plane wave pseudopotential method in Quantum Espresso [26]. For this, we have selected 163 eV kinetic energy cutoff for wavefunctions and 653 eV for charge density. The convergence threshold for selfconsistency was set to $10^{-14}$. The force convergence criterion was set to $10^{-4} eV/Å$. In the force calculations, ultrasoft pseudopotential (generated by PS library 1.00) and PBE functional were used. This method of lattice thermal conductivity calculation has been successfully used for many materials and found to be reliable [27–30]. The lattice thermal conductivity is calculated by solving linearized phonon Boltzmann equation (LBTE) [31], using the single-mode relaxation-time (SMRT) method and given by [25]

$$\kappa = \frac{1}{NV}\sum_\lambda C_\lambda \mathbf{v}_\lambda \otimes \mathbf{v}_\lambda \tau^{SMRT} \qquad (5)$$

Where V is the volume of the unit cell, $\mathbf{v}$ is the group velocity and $\tau$ is the SMRT for the phonon mode $\lambda$. The mode dependent phonon heat capacity $C_\lambda$ is defined as

$$C_\lambda = k_B \left(\frac{\hbar\omega_\lambda}{k_B T}\right)^2 \frac{\exp(\hbar\omega_\lambda/k_B T)}{[\exp(\hbar\omega_\lambda/k_B T)-1]^2} \qquad (6)$$

where $k_B$ is the Boltzmann constant and $\omega_\lambda$ is the harmonic frequency of $\lambda$ mode. The group velocity can be calculated by using the following equation:

$$v_\alpha(\lambda) = \frac{1}{\omega_\lambda}\sum_{\kappa\kappa'\beta\gamma} W_\beta(\kappa,\lambda) \frac{\partial D_{\beta\gamma}(\kappa\kappa',\mathbf{q})}{\partial q_\alpha} W_\gamma(\kappa',\lambda) \qquad (7)$$

where W is the polarization and D is the dynamical matrix. In this method, the relaxation time is considered to be equal to the phonon lifetime and given by [25]

$$\tau_\lambda^{SMRT} = \frac{1}{2\Gamma(\omega_\lambda)} \qquad (8)$$

where $\Gamma(\omega_\lambda)$ is the phonon linewidth. The Grüneisen parameter is calculated by the equation [25,32]

$$\gamma(\mathbf{q}\nu) = -\frac{V}{2[\omega(\mathbf{q}\nu)]^2}\langle e(\mathbf{q}\nu)\left|\frac{\partial D(\mathbf{q})}{\partial V}\right| e(\mathbf{q}\nu)\rangle \qquad (9)$$

where D(**q**) is the dynamical matrix, and e(**q**ν) is the phonon eigenvector at the wave vector **q**.

## 3. Result and Discussions

Fig. 1 shows the crystal structure of Cs$_2$InAgCl$_6$. The new lead free halide, Cs$_2$InAgCl$_6$, is a face-centered cubic crystal with space group $Fm\overline{3}m$ (#225) [11]. The Ag atoms occupy 4b Wyckoff; In atoms occupy 4a Wyckoff; Cs atoms occupy 8c Wyckoff; Cl atoms occupy 24e Wyckoff positions [11]. The used experimental lattice parameter is 10.467 Å [11].

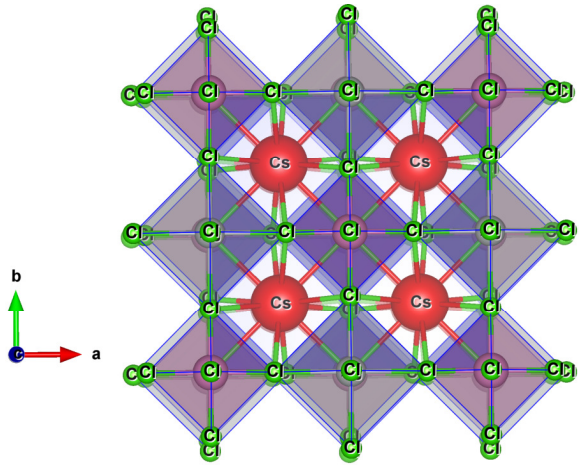

Fig. 1: Crystal structure of double perovskite Cs$_2$InAgCl$_6$.

### 3.1. *Elastic properties*

The material strength can be described by the bulk modulus, shear modulus, Young's modulus and Poisson's ratio. The well known method of calculation employed for elastic moduli and can be found in the standard articles [33–35]. The melting temperature can be predicted by the equation [36], $T_m = [553\text{K} + (5.91 \text{ K/GPa })c_{11}] \pm 300\text{K}$. The calculated elastic constants, moduli of elasticity, Poisson's ratio and melting temperature of Cs$_2$InAgCl$_6$ are presented in Table-1.

Table-1: Elastic constants, moduli of elasticity (in GPa), Poisson's ratio and melting point (in K) of $Cs_2InAgCl_6$.

| $c_{11}$ | $c_{12}$ | $c_{44}$ | B | G | E | v | B/G | $T_M \pm 300$ |
|---|---|---|---|---|---|---|---|---|
| 93.03 | 23.69 | 15.91 | 46.8 | 21.86 | 56.75 | 0.297 | 2.14 | 1102.82 |

The necessary and sufficient conditions of stability for a cubic crystal system are given as [35]:

$$C_{11} - C_{12} > 0 \; ; \; C_{11} + 2C_{12} > 0; \quad C_{44} > 0 \qquad (10)$$

The calculated elastic constants (as presented in the Table-1) indicate that $Cs_2InAgCl_6$ is elastically stable. The ductile nature can be found from both Cauchy pressure ($c_{11}$-$c_{44}$) [37] and Pugh ratio (B/G) [38]. Therefore, $Cs_2InAgCl_6$ may be used in the wide range of applications involving rapid acceleration. The shear elastic anisotropy can be calculated by [39], $A = \frac{2C_{44}}{C_{11}-C_{12}}$. The value of $A$ is 0.46 that indicates the anisotropic nature of $Cs_2InAgCl_6$.

The Vickers hardness can be calculated by using Pugh ratio (G/B) in the equation [40], $H_V = 2\left(\left(\frac{G}{B}\right)^2 G\right)^{0.585} - 3$. Our calculated Vickers hardness is 1.98 and indicates that $Cs_2InAgCl_6$ is relatively low hard. The Debye temperature is related to thermal stability and can found by the equation [41]

$$\theta_D = \frac{h}{k_B}\left(\frac{3N}{4\pi V}\right)^{1/3}\left[\frac{1}{3}\left(\frac{2}{v_t^3}+\frac{1}{v_l^3}\right)\right]^{-\frac{1}{3}} \qquad (11)$$

where, $v_l$ and $v_t$ are given by $v_l = \left(\frac{3B+4G}{3\rho}\right)^{1/2}$, $v_t = \left(\frac{G}{\rho}\right)^{1/2}$. The calculated Debye temperature is 253K and this low value indicates that lattice thermal conductivity of $Cs_2InAgCl_6$ should be small.

## 3.2. *Electronic properties*

The calculated band structure of $Cs_2InAgCl_6$ is shown in the Fig. 2. Some energy bands are dispersive and others are nondispersive. The dispersive conduction bands with a typical bandwidth ~5.46 eV in mBJ but ~3.46 eV in PBE which is consistent with other calculation [12] and arise from In-5s states due to the delocalized nature of these sates. The maximum peak of the valence band at $\Gamma$ and X-points mainly comes from Ag-4d and Cl-3p states. It is found that the conduction bands due to Ag-5s states have higher energy (appeared only in PBE band structure, black line) than these for In-5s states and this leads to another conduction state at L-point. The delocalized nature of In-5s states and inversion symmetric nature leads to the even parity of maximum conduction band at $\Gamma$. However, bandgap energy state at the $\Gamma$-point has the highest energy and thus also leads to even parity. Therefore, transitions forbidden by parity conservation should be observed. The conduction bands are less flat than valence bands. These flat bands arise from the strong hybridization of Cl-3p, Ag-4d and In-4d orbitals as clearly can be seen from the projected density of states (PDOS) shown in the Fig. 3. The fundamental direct bandgap is 1.817 eV, calculated by using PBE functional which is much larger than the calculated bandgap [12] 1.03 eV by using same functional. Our PBE calculation underestimates the bandgap by 1.483 eV. The modified TB-mBJ calculation overestimates the bandgap by 0.37 eV. However, TB-mBJ calculated fundamental bandgap of $Cs_2InAgCl_6$ is reasonably close to the experimentally measured optical bandgap 3.3 eV [11].

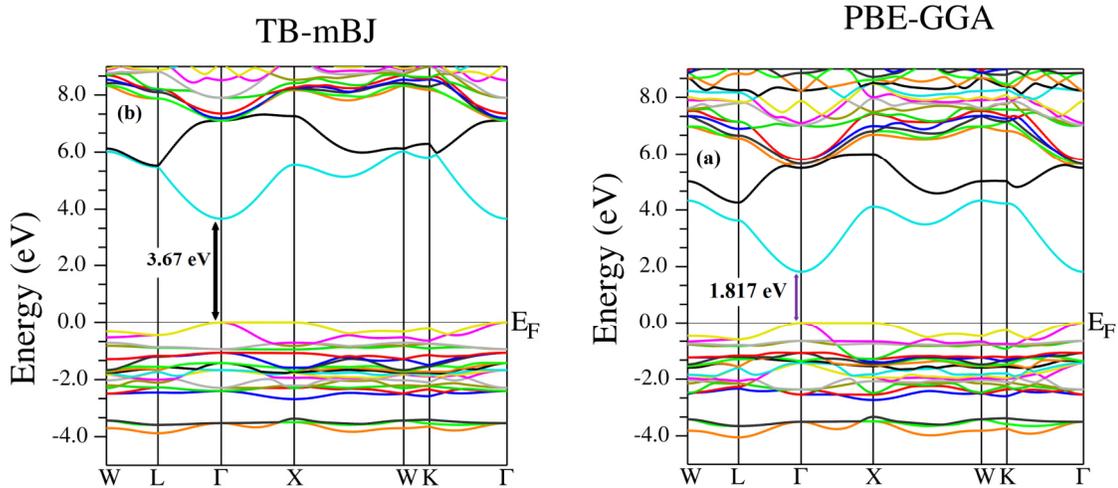

Fig. 2: Band structure of double perovskite $Cs_2InAgCl_6$ by using: (a) PBE functional, (b) TB-mBJ functional.

The fundamental direct bandgap is 1.817 eV, calculated by using PBE functional which is much larger than the calculated bandgap [12] 1.03 eV by using same functional. Our PBE calculation underestimates the bandgap by 1.483 eV. The modified TB-mBJ calculation overestimates the bandgap by 0.37 eV. However, TB-mBJ calculated fundamental bandgap of $Cs_2InAgCl_6$ is reasonably close to the experimentally measured optical bandgap 3.3 eV [11]. A comparison of experimentally measured bandgap and theoretically calculated bandgap by using different functionals are presented in Table-2. Thus, it is clear that the modified TB-mBJ functional can provide a correct bandgap of semiconductors. The highest peak (around -6 eV) in the DOS comes from the sigma bonding combinations of Cs-5p, In-5s, and In-5p orbitals. The second peak (around -0.5 eV) arises from the strong hybridization of Ag-4d, In-4d, and Cl-3p orbitals. It is clear from the Fig. 3 that TB-mBJ functional calculation shifts the energy level of density of states (DOS) as expected (see band structure).

Table-2: Comparison of the measured and calculated bandgap of $Cs_2InAgCl_6$.

| Method | Bandgap (eV) | Ref. |
| --- | --- | --- |
| EXP-I | 3.2 | [11] |
| EXP-II | 3.3 | [13] |
| PBE0 | 2.7 | [11] |
| HSE | 2.6 | [11] |
| PBE | 1.03 | [12] |
| PBE-1 | 1.817 | This |
| TB-mBJ | 3.67 | This |

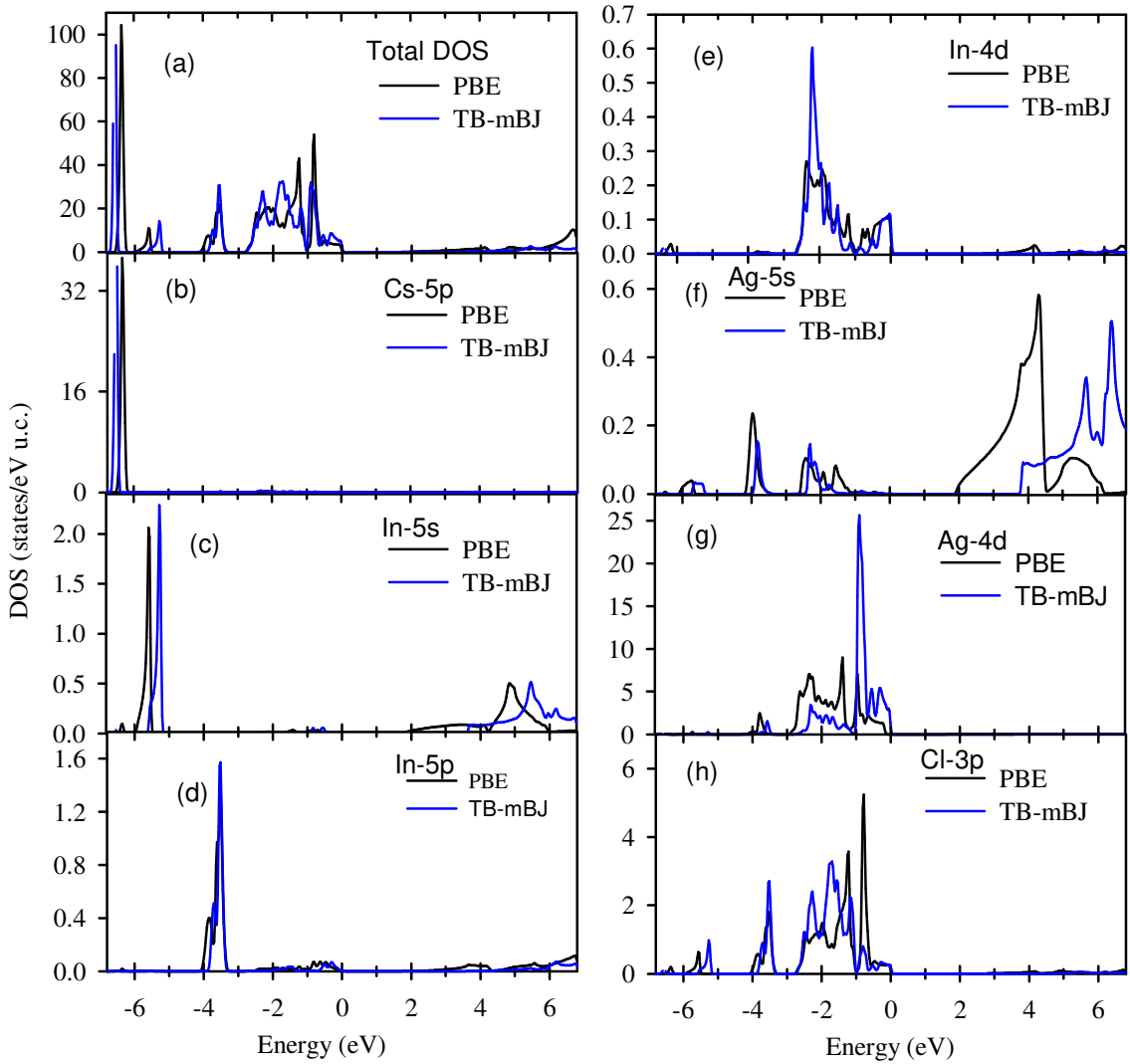

Fig. 3: Total and projected density of states of $Cs_2InAgCl_6$. The zero energy indicates the Fermi level.

## 3.1 Thermoelectric properties

The calculated phonon group velocity is presented in the Fig. 4. The phonon group velocity is quite low as expected (since Debye temperature is very low). The large weighted phase space may be responsible for such low phonon group velocity.

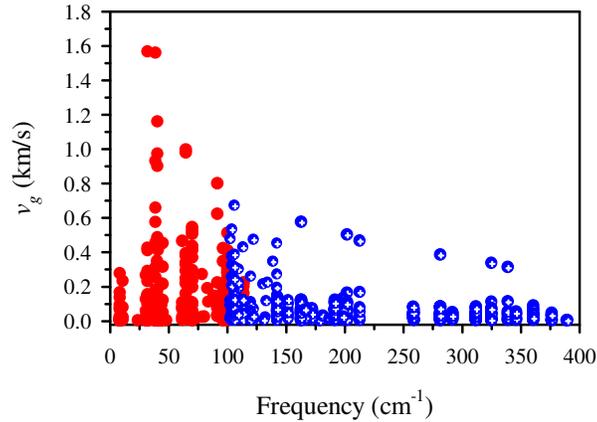

Fig.4: Variation of phonon group velocity with frequency. The red circle indicates the acoustic mode, and blue circle (with cross) indicates optical mode.

To understand the anharmonicity in $Cs_2BiAgCl_6$ crystal, the phonon Grüneisen parameter is plotted with frequency in the Fig. 5. The larger value of Grüneisen parameter indicates high anharmonicity in the crystal and hence large phonon scattering. Our calculated phonon Grüneisen parameter $Cs_2BiAgCl_6$ is very large and indicates the large phonon scattering in the $Cs_2BiAgCl_6$ crystal. Such intrinsic large phonon scattering leads to large weighted phase space and hence low phonon group velocity giving arise the quite low lattice thermal conductivity in $Cs_2BiAgCl_6$. The calculated relaxation time from phonon linewidths is presented in the Fig. 6. The relaxation time decreases with temperature as expected. It is clear that phonon-relaxation is large and 0.11 ps at 300 K. The calculated lattice thermal conductivity by using relaxation time approximation (RTA) is presented in the Fig. 6(b). The calculated lattice thermal conductivity is 0.2 W/m K at 300 K. Our calculated lattice thermal conductivity of $Cs_2BiAgCl_6$ is slightly larger than that obtained in $As_2Se_3$ (0.14 W/m K) [42].

The quite low phonon group velocity, and large phonon scattering give rise to such low lattice thermal conductivity.

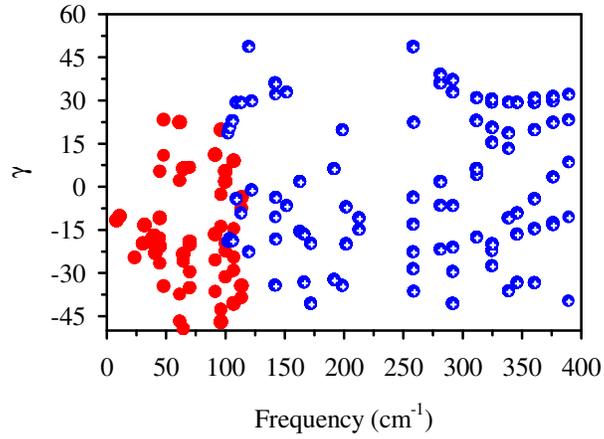

Fig. 5. Variation of phonon Grüneisen parameter with frequency. The red circle indicates the acoustic mode, and blue circle (with cross) indicates optical mode.

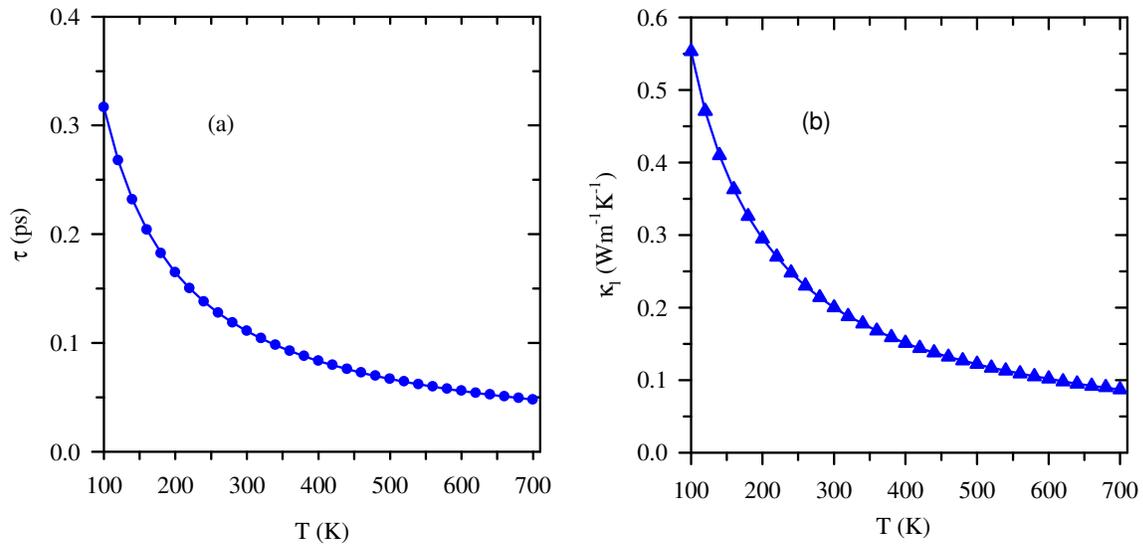

Fig. 6: Temperature dependence of (a) Relaxation time and (b) lattice thermal conductivity of $Cs_2BiAgCl_6$.

The calculated Seebeck coefficient of $Cs_2BiAgCl_6$ sharply decreases with temperature. This implies that electron and hole contributions to the transport properties increase with temperature that leads to shifting the Fermi level in the middle of the bandgap. Although this is usually observed for n-type materials, $Cs_2BiAgCl_6$ is found to be p-type material as $S$ remains positive. The small value of Fermi energy 0.1847 eV leads to more electrons occupy the lower energy states and hence may be responsible for such kind of behavior. The Seebeck coefficient calculated by using TB-mBJ potential is larger than that by PBE potential upto 300K but from 300K to 360K, both functionals provide almost the same Seebeck coefficient. However, after 380K the Seebeck coefficient obtained by TB-mBJ more sharply decreases with temperature than that by PBE (see Fig. 7(a)). The room temperature Seebeck coefficient is ~199 $\mu$V/K for both PBE and TB-mBJ potentials. Large bandgap and flat conduction bands are responsible for such high Seebeck coefficient. The electrical conductivity ($\sigma/\tau$) increases with the temperature implies the semiconducting nature of $Cs_2BiAgCl_6$ as shown in the Fig. 7(b). Below 360 K, the electrical conductivity calculated by PBE functional is same to that calculated by TB-mBJ. After this temperature, the difference between them increases with temperature. The calculated electrical conductivity ($\sigma$) at 300K are $7.81 \times 10^5$ and $7.33 \times 10^5$ S/m using PBE and TB-mBJ potentials, respectively. The high electrical conductivity of $Cs_2BiAgCl_6$ arises from narrow electron conduction band causing the large relaxation time. The electronic part of the thermal conductivity ($\kappa/\tau$) also increases with temperature. The electronic thermal conductivity increases similar to the electrical conductivity. Therefore, it is preferable to increase Seebeck coefficient by doping with the suitable element. The calculated electronic part of thermal conductivity at 300K is 12.83 and 11.99 W/mK by PBE and TB-mBJ potentials, respectively. The total thermal conductivity increases with temperature as expected (shown in the Fig. 7 (d)). The calculated total thermal conductivity at 300K 13.03 and 12.18 W/mK by PBE and TB-mBJ, respectively. The power factor ($S^2\sigma$) of $Cs_2InAgCl_6$ decreases with temperature for both TB-mBJ and PBE functionals as shown the Fig. 8(a).

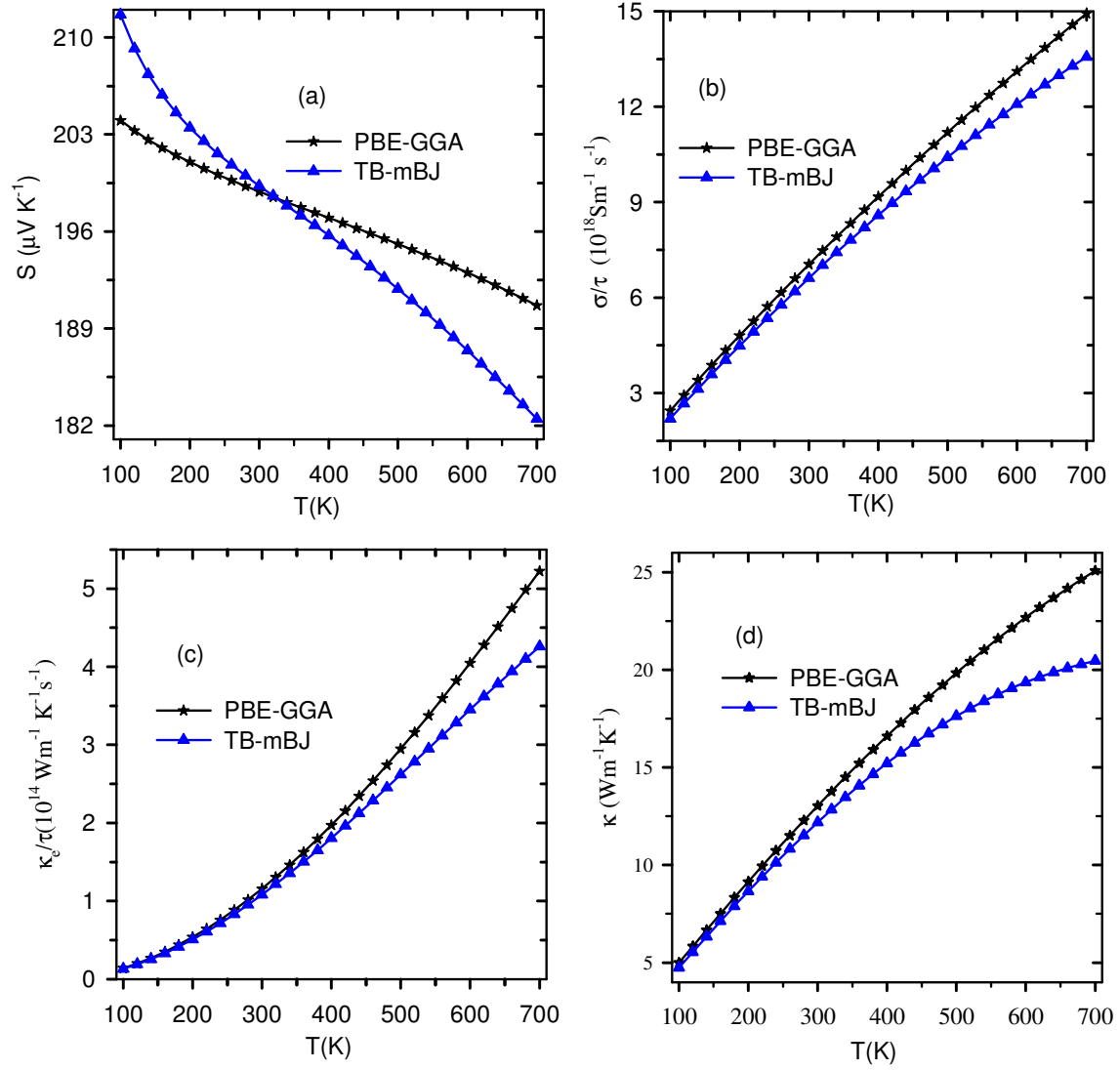

Fig.7: Temperature dependence of thermoelectric transport properties of $Cs_2InAgCl_6$: (a) Seebeck coefficient (S), (b) electrical conductivity ($\sigma/\tau$), (c) electronic part of the thermal conductivity ($\kappa_e/\tau$), and (d) total thermal conductivity ($\kappa$).

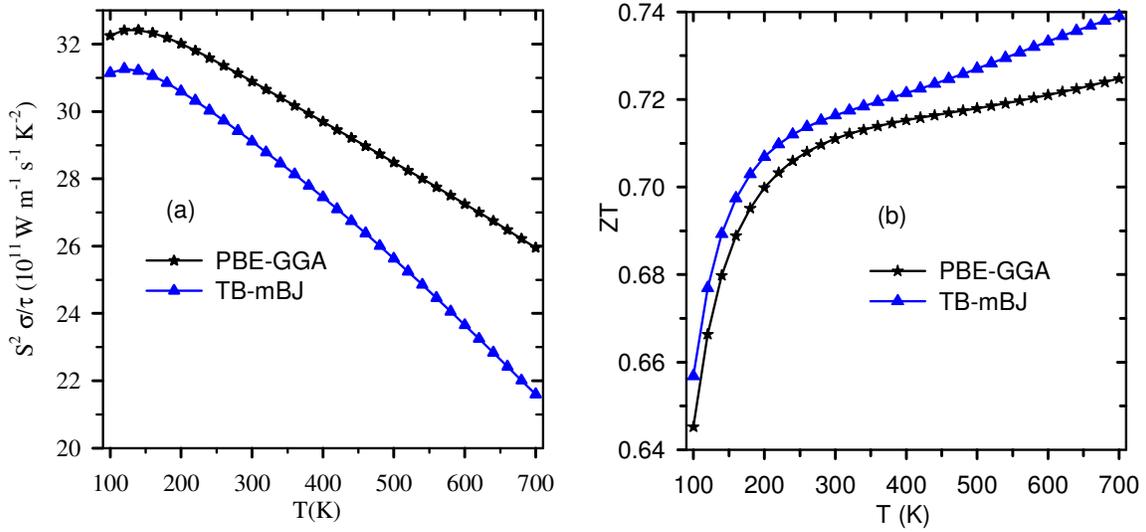

Fig. 8: Temperature dependence of (a) Power factor ($S^2\sigma$) and (b) thermoelectric figure of merit of $Cs_2InAgCl_6$.

The calculated room temperature (300 K) power factor is ~31 and ~29 mW/m $K^2$ by PBE and TB-mBJ potentials, respectively. The temperature dependence of thermoelectric figure of merit is illustrated in Fig. 8(b). It is obvious from Fig. 8(b) that ZT is little higher in TB-mBJ calculation than PBE because thermal conductivity is lower in TB-mBJ calculation than PBE. The calculated figure of merit by TB-mBJ potential at 300 and 700K are 0.72 and 0.74, respectively. The obtained results from first principles calculation reveal that $Cs_2InAgCl_6$ is a promising thermoelectric material.

## 4. Conclusions

In summary, we have performed DFT (density functional theory) calculations to study the elastic and electronic properties of indium-based double-perovskite halide $Cs_2InAgCl_6$ and solved Boltzmann transport equation to study the thermoelectric transport properties. We have found $Cs_2InAgCl_6$ to be elastically stable, ductile and relatively low hard material. Our calculated direct bandgap 3.67 eV (by TB-mBJ functional) fairly agrees with the experimentally measured value 3.3 eV but PBE functional underestimates the bandgap by 1.483 eV. We have calculated the relaxation time and lattice thermal conductivity by using relaxation time approximation (RTA) within the supercell approach. The lattice thermal conductivity ($\kappa_l$) is quite low (0.2 Wm$^{-1}$K$^{-1}$). The low phonon group velocity in the large weighted phase space and anisotropic nature of the material are responsible for small $\kappa_l$. The room temperature Seebeck coefficient using PBE and TB-mBJ potentials exhibit the same value ~199 $\mu$VK$^{-1}$. Such high thermopower (Seebeck coefficient) arises from the combination of the flat conduction band and large bandgap. We obtain power factors at 300K by using PBE and TB-mBJ potentials are ~29 and ~31 mWm$^{-1}$K$^{-2}$, respectively and the corresponding thermoelectric figure of merit of Cs$_2$BiAgCl$_6$ are 0.71 and 0.72. However, the maximum ZT value obtained at 700K is ~0.74 for TB-mBJ potential. The exciting results will inspire further experimental study on thermoelectric properties of Cs$_2$InAgCl$_6$ making it a promising material for thermoelectric device applications.